\documentclass[twocolumn,showpacs,amsmath,amssymb,superscriptaddress,prl]{revtex4}
\usepackage{graphicx}
\usepackage{dcolumn}
\usepackage{bm}
\usepackage{indentfirst}
\begin{document}
%
%
\title{Critical and excess current through an open quantum dot:\\
Temperature and magnetic field dependence}
\author{H. Ingerslev J\o rgensen}
\email{hij@fys.ku.dk}
\affiliation{Nano-Science Center, Niels Bohr
Institute, University of Copenhagen, Universitetsparken 5,
DK-2100~Copenhagen \O , Denmark}
\author{K. Grove-Rasmussen}
\altaffiliation{Present address: NTT Basic Research Laboratories,
3-1 Morinosato Wakamiya, Atsugi-shi, 243-0198 Kanagawa, Japan}
\affiliation{Nano-Science Center, Niels Bohr Institute, University
of Copenhagen, Universitetsparken 5, DK-2100~Copenhagen \O ,
Denmark}
\author{K. Flensberg}
\affiliation{Nano-Science Center, Niels Bohr Institute, University
of Copenhagen, Universitetsparken 5, DK-2100~Copenhagen \O ,
Denmark}
\author{P. E. Lindelof}
\affiliation{Nano-Science Center, Niels Bohr Institute, University
of Copenhagen, Universitetsparken 5, DK-2100~Copenhagen \O ,
Denmark}
\date{\today}
\begin{abstract}
We present measurements of temperature and magnetic field dependence
of the critical current and excess current in a carbon nanotube
Josephson quantum dot junction. The junction is fabricated in a
controlled environment which allows for extraction of the full
critical current. The measurements are performed in the open quantum
dot regime, and fitted to theory with good qualitative agreement. We
also show how to extract level spacing, level broadening, and
charging energy of an open quantum dot from a bias spectroscopy
plot.
\end{abstract}
\maketitle
\begin{figure}
\begin{center}
\includegraphics[width=0.45\textwidth]{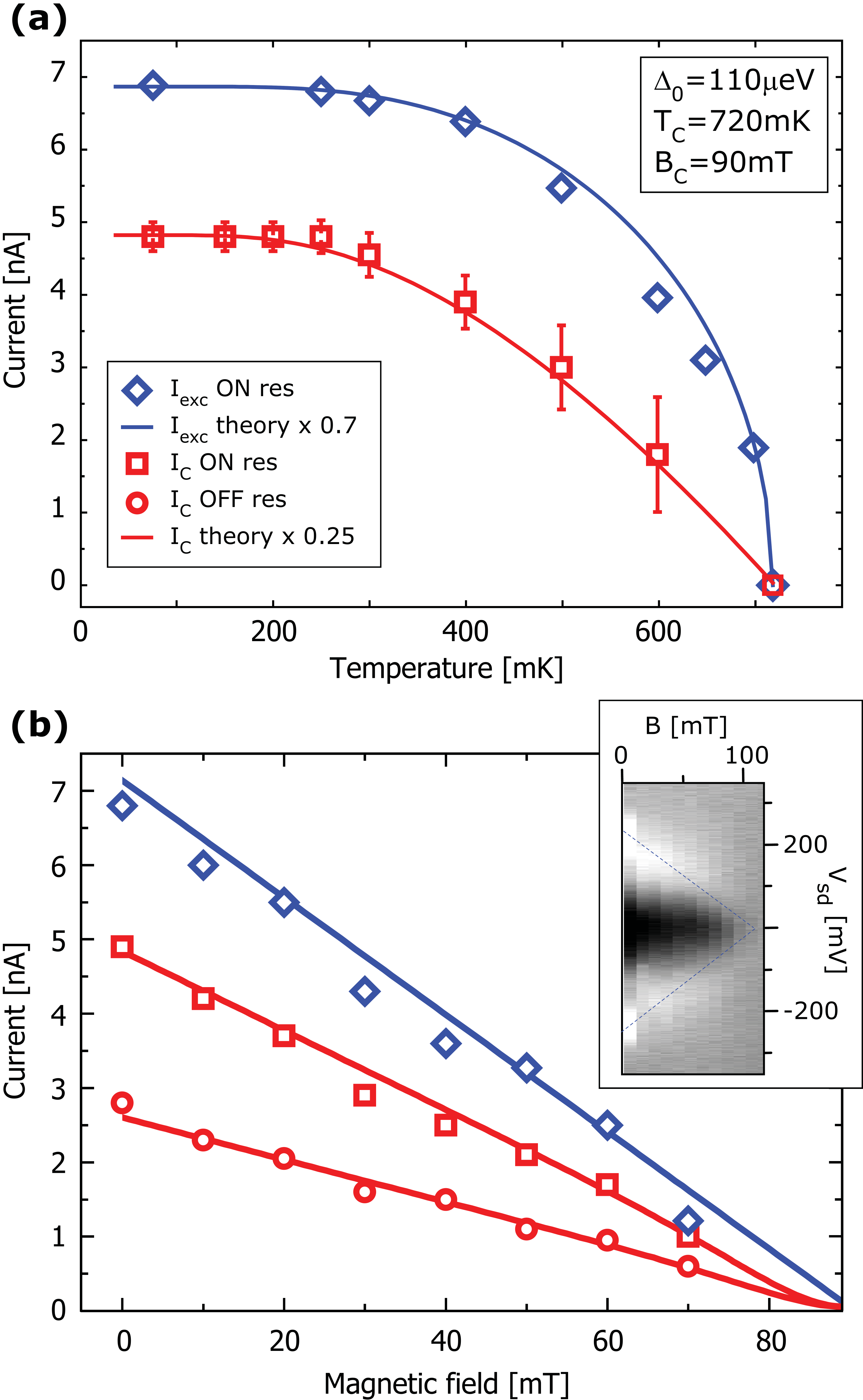}
\end{center}
\caption{(color online) (a) Temperature dependence of the measured
critical current (squares) and excess current (diamonds) on
resonance. (b) Magnetic field dependence of the measured critical
current on and off resonance (squares and circles), and excess
current on resonance (diamonds). The normal state zero-bias
conductance is 2.6\,$e^2/h$ on resonance and 1.4\,$e^2/h$ off
resonance. The solid lines in both (a) and (b) are the predicted
curves for a superconducting quantum point contact multiplied by a
constant factor of 0.25 for the critical current and 0.7 for the
excess current. Insert shows the magnetic field dependence of the
sub gap structure of a similar device in the Coulomb blockade
regime.} \label{fig:4}
\end{figure}

Nanoscale Josephson quantum dot junctions are intriguing devices
showing several interesting physical phenomena. Supercurrent,
Andreev reflections, quasiparticle transport, and excess current
have all been studied in junctions where a nanotube or nanowire
constitute the quantum dot
\cite{kasumov,yong-joo,takesue,pablo,jorgensen,Pallecchi,zhang}.
Furthermore, the interplay between these Josephson junction related
phenomena and correlations as the Kondo effect
\cite{buitelaar_mar,buitelaar_s_kondo,grove_njp,sand,eichler_design}
and the 0-$\pi$ transition for more weakly coupled junctions has
been explored  \cite{jordan,Cleuziou,jorgensen_nanolett}.
\newline \indent
In this paper, we present experimental results in the strongly
coupled regime for a Josephson quantum dot junction realized in a
carbon-nanotube. Inspired by Ref.[\onlinecite{steinbach}], we
utilize a designed external circuit in order to control the phase
fluctuations which enables us to infer the true magnitude of the
critical current, $I_C$, from the measurable critical
current/switching current, $I_m$, by a fitting procedure
\cite{jorgensen_nanolett}. $I_m$ can significantly differ from $I_C$
as demonstrated previously for nanotube-based Josephson junctions
\cite{pablo,jorgensen,jorgensen_nanolett}. Here we analyze the
magnetic field dependence and temperature dependence of both the
critical current and excess current.
\begin{figure}
\begin{center}
\includegraphics[width=0.48\textwidth]{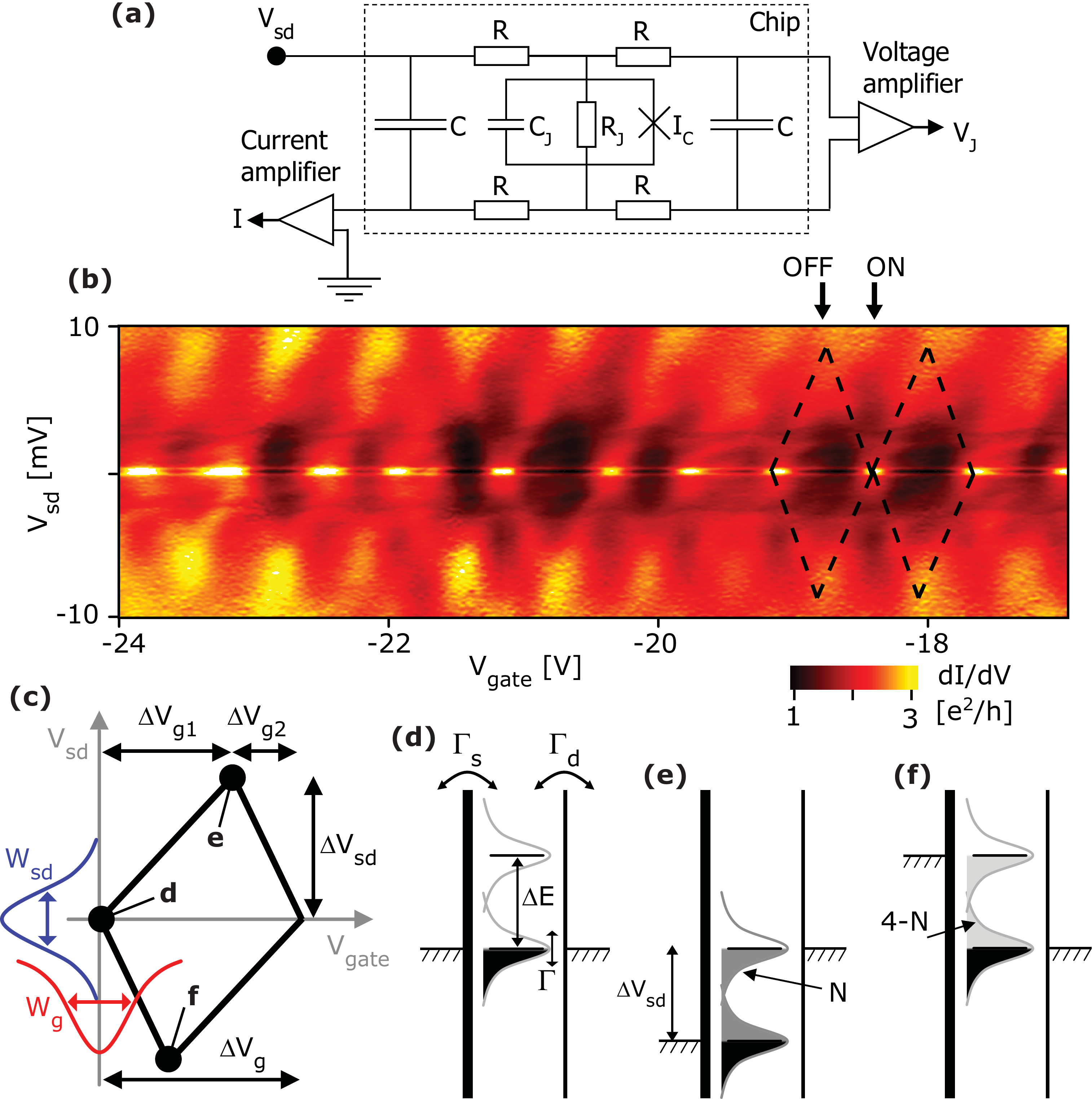}
\end{center}
\caption{(color online) (a) Inside the dashed square: Schematic
circuit of on-chip components of the Josephson junction. Outside
dashed square: Four probe voltage bias setup for measuring junction
voltage $V_J$ vs. current $I$. The full Josephson junction consists
of both the superconductor-nanotube-superconductor junction,
represented by a Josephson element $I_C$ in parallel with a junction
capacitor $C_J$ and junction resistor $R_J$, and on-chip resistors
$R$ and capacitances $C$. (b) Bias spectroscopy plot of differential
conductance versus source-drain, and gate voltage. (c) Schematic of
a Fabry-Perot diamond. (d-f) Schematic transport diagrams at
zero-bias resonance (d), positive-bias resonance (e), and
negative-bias resonance (f).} \label{fig:1}
\end{figure}
\newline \indent
The devices are fabricated on a degenerately doped silicon wafer
with a 0.5\,$\mu$m layer of SiO$_2$. Carbon nanotubes are grown from
islands of catalyst material and contacted by small electrodes of
superconducting trilayers of 5nm Ti, 60nm Al and 5nm Ti. The
superconducting electrodes are kept small to reduce junction
capacitance. Each superconducting electrode is contacted by two
normal metal leads to bonding pads which enables four probe
measurements. The measurements are performed in a $^3$He-$^4$He
dilution fridge with a base electron temperature of 75\,mK. Inside
the dashed square in Fig.\,\ref{fig:1}(a) we show a schematic
circuit diagram of the on-chip components of the full Josephson
junction. The fabrication is similar to
Ref.\,[\onlinecite{jorgensen_nanolett}].
The superconductor-nanotube-superconductor junction is represented
by a Josephson element (cross), a junction resistor $R_J$, and a
junction capacitor $C_J$.
The Josephson element has a current-phase relation, which we in the
fitting procedure (see below) assume to be $I(\phi)=I_C \sin
(\phi)$, with $\phi$ being the phase difference between the two
superconducting and $I_C$ the critical current. However, the
sinusoidal form of this relation is not in general true and this may
cause some inaccuracy in the determination of the critical current.
In Ref.\,[\onlinecite{jorgensen_nanolett}] we show that the
difference between the extracted $I_C$ using either $\sin(\phi)$ or
the correct functional form of the current-phase relation is in fact
small and moreover largest near the $0$-$\pi$ transition relevant
only for closed dots. We therefore expect the simpler relation also
to be a reasonable approximation in the case of open dots (which
allows us to use the Ivanchenko-Zil'berman relation in
Eq.\,\eqref{eq:1}). At sub-gap bias voltages $R_J$ accounts for
current due to multiple Andreev reflections and at higher bias
voltages it accounts for quasi-particle transport. The capacitance
between the superconducting electrodes ($C_J \sim 5$\,fF), and
between bonding pads ($C \sim 1$\,pF) is estimated as a parallel
plate capacitance through the back gate. We have fabricated long
thin metal leads with a measured resistance of $R \sim 1$\,k$\Omega$
between the bonding pads and the superconducting electrodes which,
as will be shown later, is crucial for increasing the measurable
critical current.
\newline \indent
In Fig.\,\ref{fig:1}(b) we show a bias spectroscopy plot of
differential conductance versus source-drain voltage ($V_{sd}$), and
gate voltage ($V_{gate}$). Regular conductance oscillations in both
source-drain and gate voltage is seen due to tuning of successive
energy levels in the dot, with a separation (level spacing) $\Delta
E$, on and off resonance. Bias spectroscopy plots with the leads in
the normal state ($B=150$\,mT) (not shown) show conductances at the
resonances ranging from 2 to 3.5\,$e^2/h$. That high conductance is
only allowed when the degeneracy of each energy level is four-fold
(spin and orbital), and when the broadening of each energy level
$\Gamma = \Gamma_s + \Gamma_d$, where the $\Gamma_s/\hbar$
($\Gamma_d/\hbar$) is the tunnel rate through the source (drain)
barrier, is larger than the Coulomb repulsion energy for adding an
electron to the dot (charging energy) $U_C=e^2/C$. This regime
($\Delta E
> \Gamma > U_C$) is often called the Fabry-Perot regime
\cite{Liang,pablo}, and the dot is termed an open quantum dot. We
will now analyze the bias spectrum and extract energy parameters,
tunnel couplings, and capacitances.
\newline\indent
From the size of the Fabry-Perot diamond we have the following three
equations, where we apply the source-drain voltage to the source
electrode and keep the drain electrode at ground (see
Fig.\,\ref{fig:1}(c-f)).
\begin{eqnarray}
e\, \Delta V_{sd} & = & \Delta E , \\
e\frac{C_g}{C}\Delta V_g & = & \Delta E + 4 U_C , \\
e\frac{C_g}{C}\Delta V_{g1} + e\frac{C_s}{C}\Delta V_{sd} & = &
\Delta E + N\, U_C .
\end{eqnarray}
Where $e$ is the electron charge, $\Delta V_g$, $\Delta V_{g1}$, and
$\Delta V_{sd}$ determines the the size of the Fabry-Perot diamond
as shown in Fig.\,\ref{fig:1}(c), and $C_g$, $C_s$, $C_d$, and $C$
are the capacitance of the dot to gate, source, drain and the total
capacitance. $N$ is the equilibrium number of electrons added to the
dot from a zero-bias resonance to the first positive-bias resonance,
i.e., from position d to e in Fig.\,\ref{fig:1}. $N$ can be given in
terms for of the tunnel barrier asymmetry
($\alpha=\Gamma_s/\Gamma_d$):
\begin{equation}
N = 4\,\frac{\Gamma_d}{\Gamma} = \frac{4}{\alpha +1} ,
\end{equation}
$\alpha$ can be found from the conductance at resonance: $G_0 =
16\alpha/(\alpha + 1)^2 e^2/h$. For the resonance indicated with an
arrow in Fig.\ref{fig:1}(b) we find $G_0\sim 2.6e^2/h$, $\alpha \sim
0.3$, and $N\sim 3.1$.
\newline\indent
From the width (full width at half maximum) at resonances in gate
($W_g$) and bias ($W_{sd}$) (see Fig.\ref{fig:1}(c)) we can set up
the following two equations:
\begin{eqnarray}
e\frac{C_g}{C}W_g & = & \Gamma + N^\prime \, U_C , \\
e\,W_{sd} & \approx & 2\Gamma ,
\end{eqnarray}
where the second equation is a good approximation when the asymmetry
of the capacitive or tunnel coupling is not too large (see
appendix). $N^{\prime}$ is the number of electrons added to the dot
between $V_{gate}=\pm W_g/2$ from resonance. We estimate
$N^{\prime}$ by integrating a Lorentzian density of state for each
energy level on the dot:
\begin{equation}
N^\prime = \int_{-W_g/2}^{W_g/2} \sum_j
\frac{4}{\pi}\,\frac{\frac{1}{2}W_g}{\left(\epsilon +j\Delta
V_g\right)^2 + \left(\frac{1}{2}W_g\right)^2} \, d\epsilon
\end{equation}
where the sum should include an appropriate number of energy levels.
If only one energy level is included ($j=0$) $N^{\prime}=2$, but for
increasing number energy levels included $N^{\prime}$ saturates at a
higher number (since the tails of the other levels contribute). For
the device analyzed in paper it saturates at $N^{\prime} \sim 2.5$.
\newline \indent
By solving the equations above we can find expressions for the
following parameters:
\begin{equation}\label{eq:parameters}
\begin{array}{lclcl}
\Delta E & = & e\,\Delta V_{sd} & \sim & 9\,\textrm{meV} \\
U_C & = & \frac{W_g \Delta E - e\Delta
V_g\,W_{sd}\frac{1}{2}}{N^\prime
\Delta V_g - 4W_g} & \sim & 0.5\,\textrm{meV} \\
C_g & = & \left( \frac{\Delta E}{U_C} +4 \right)\frac{e}{\Delta V_g} & \sim & 4.3\, \textrm{aF} \\
C_s & = & \frac{e\,C_g\Delta V_{g2} - (4-N)e^2}{\Delta E} & \sim & 152\, \textrm{aF}\\
C_d & = & \frac{e^2}{U_C} - C_s - C_g & \sim & 158\, \textrm{aF} \\
C & = & C_s + C_d + C_g  & \sim & 315\, \textrm{aF} \\
\Gamma & = & \frac{W_{sd}}{2}\, e & \sim & 4.3\, \textrm{meV} \\
\Gamma_s & = & \Gamma\,\frac{\alpha}{\alpha + 1} & \sim & 1\, \textrm{meV} \\
\Gamma_d & = & \Gamma\,\frac{1}{\alpha + 1} & \sim & 3.3\,
\textrm{meV}
\end{array}
\end{equation}
We have in the right hand column estimated the parameters for the
device analyzed in this paper \footnote{$\Delta V_g=0.8$\,V, $\Delta
V_{g1}=0.45$\,V, $\Delta V_{g2}=0.35$\,V, $\Delta V_{sd}=9$\,mV,
$W_g=0.4$\,V, $W_{sd}=8.5$\,mV}. Note that $\Delta E
> \Gamma
> U_C
> \Delta_0$, where $\Delta_0 \sim 0.11$\,meV is the superconducting
energy  gap (see below).
\begin{figure}
\begin{center}
\includegraphics[width=0.48\textwidth]{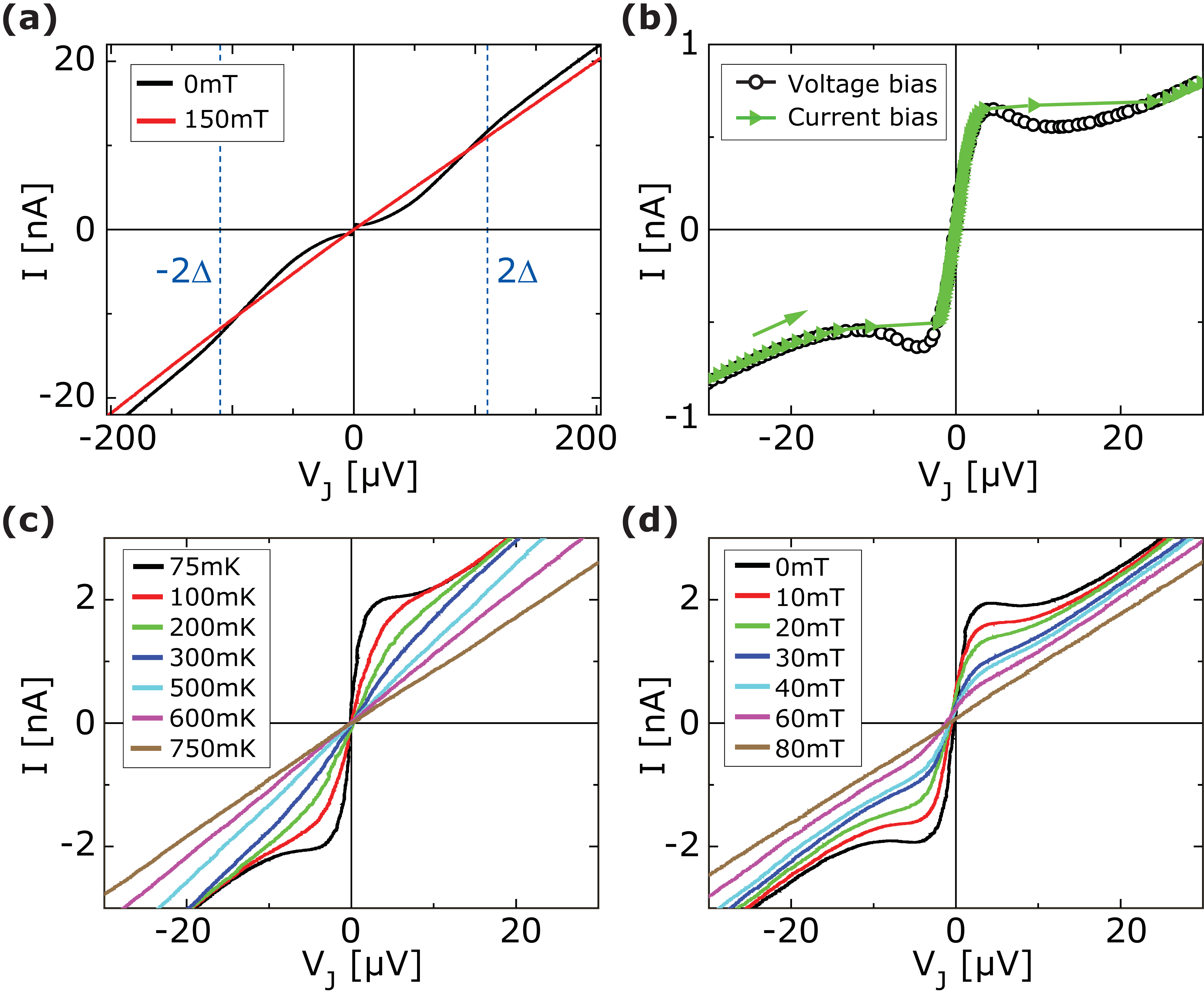}
\end{center}
\caption{(color online) Current versus junction voltage on (a-b) and
off (c-d) resonance at positions indicated in Fig.\,\ref{fig:1}(b).
(a) Black curve is with the electrodes in the superconducting state,
and red curve with a small magnetic field (150\,mT) applied to
suppress the superconductivity. (b) Close-up of the supercurrent
branch from (a), measured with a voltage bias setup (circles) and a
current bias setup (triangles). (c-d) Dependence of the diffusive
supercurrent branch on temperature (c) and magnetic field (d).}
\label{fig:2}
\end{figure}
\newline \indent
We now return to the measurements shown in Fig.\,\ref{fig:1}(b),
where two parallel conductance ridges are observed at low bias due
to the density of states in the superconducting electrodes. The
separation between these two rides is $4\Delta_0/e$, yielding
$\Delta_0 \sim 0.11$\,meV. In the following we focus on measurements
performed on and off zero-bias resonance at the two indicated
positions in Fig.\ref{fig:1}(b). Current versus junction voltage
($IV_J$ curves) off resonance for large scale voltages is shown in
Fig.\,\ref{fig:2}(a), where the black curve is with superconducting
electrodes and the red curve is with a small magnetic field
(150\,mT) to suppress the superconductivity. At high bias $V_{sd} >
2\Delta_0/e$ transport is governed by quasiparticle transport and
one Andreev reflection processes yielding an excess current, while
at sub-gap bias $V_{sd} < 2\Delta_0/e$ transport are governed by
Andreev reflections and supercurrent \cite{jorgensen}. A close-up at
very low bias voltages, shown in Fig.\,\ref{fig:2}(b), reveals a
pronounced supercurrent branch with finite resistance, a so-called
diffusive supercurrent branch \cite{jorgensen_nanolett}. The black
circles are measured with a voltage bias setup as shown in
Fig.\,\ref{fig:1}(a), while the green triangles are measured with a
current bias setup (sweeping from negative to positive current). For
voltage bias measurements we have observed no hysteresis or
switching in the $IV_J$ curves at any gate voltages. But for current
bias measurements switching and hysteresis are observed whenever the
full $IV_J$-curve has local minima and maxima, as observed in
Fig.\,\ref{fig:2}(b). Such local minima and maxima will for current
bias measurements lead to switching in voltage and result in a
hysteretic $IV_J$-curve. To resolve the full $IV_J$-curve we have
therefore used voltage bias measurements in this paper.
\begin{figure}
\begin{center}
\includegraphics[width=0.48\textwidth]{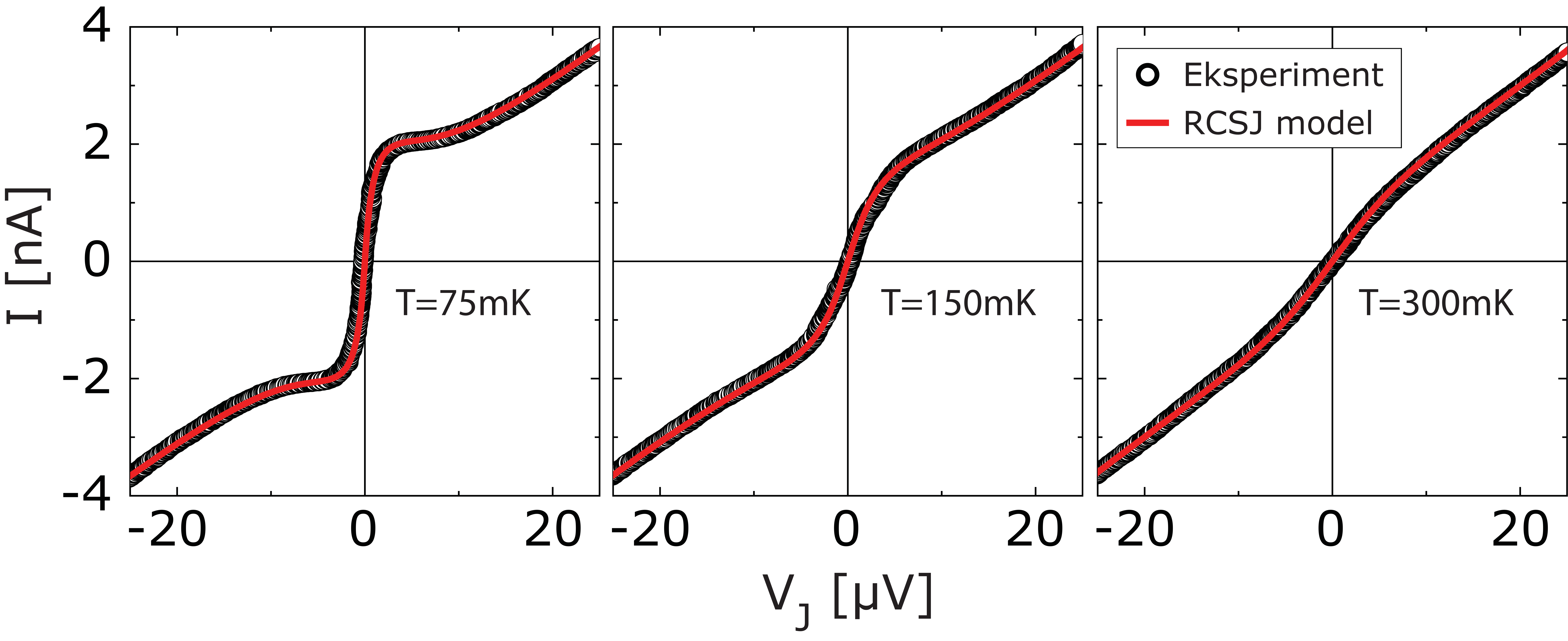}
\end{center}
\caption{(color online) Current versus junction voltage on resonance
at position indicated in Fig.\,\ref{fig:1}(b) for three different
temperatures. From left to right: 75\,mK, 150\,mK, and 300\,mK. Four
probe voltage bias measurement (circles), and fit (solid red line)
using Eq.\,\eqref{eq:1} with $R_J=7.7k\Omega$, $R=1k\Omega$, the
temperature at which the curve is measured, and $I_C = 4.8$\,nA,
4.8\,nA, and 4.6\,nA from left to right.} \label{fig:3}
\end{figure}
\newline \indent
In Fig.\,\ref{fig:2}(c) and (d) we show the temperature and magnetic
field dependence of the diffusive supercurrent branch, which we will
analyze in the following. The zero bias slope of the diffusive
supercurrent branch in Fig.\,\ref{fig:2} yields a resistance of the
order kilo ohm. For a Josephson quantum dot junction with only two
channels as for a nanotube the Josephson energy $E_J = \hbar I_C /
2e$ can be comparable to the temperature of the cryostat. Thermal
fluctuations will therefore lead to fluctuations in the phase
difference across the junction, and consequently give a supercurrent
branch with finite resistance. In order to dampen these phase
fluctuations and thereby increase the size of the supercurrent
branch, we have designed the environment of the
superconductor-nanotube-superconductor junction as described in
Ref.\,[\onlinecite{jorgensen_nanolett}]. The quality factor for the
junction is $Q<0.5$, i.e., strongly damped. The full $IV_J$-curve
for a damped Josephson junction including the external components
(without $R_J$) was calculated by Ivanchenko and
Zil'berman\cite{ivan-zil} and used with great success by Steinbach
et.~al.~\cite{steinbach}. Since this device has considerable current
contribution from multiple Andreev reflections at sub-gap bias
voltage we have to a rough approximation included a constant
resistor $R_J$. The full $IV_J$-curve can then be calculated
as\cite{jorgensen_nanolett}
\begin{equation}\label{eq:1}
I(V_{sd})=I_C Im \left( \frac{I_{1-\eta i}(E_J/k_B T)}{I_{-\eta
i}(E_J/k_B T)} \right) + \frac{V_J}{R_J}
\end{equation}
where $I_n(x)$ is the modified Bessel function of complex order, and
$\eta = (\hbar V_{sd})/(2eRk_B T)$. To plot $I(V_{sd})$ versus $V_J$
instead of $V_{sd}$ we can use that $V_J = V_{sd} - RI(V_{sd})$.
There are two fitting parameters in this theory, the temperature
dependent critical current $I_C(T)$ and $R_J$. In Fig.\,\ref{fig:3}
we show three I versus $V_J$ curves measured at the same gate
voltage for increasing temperatures. From left to right: 75\,mK,
150\,mK, and 300\,mK. The black circles are the measurement and the
solid red curve is theoretical fit with Eq.\,\eqref{eq:1}. The three
fits are made with $R_J=7.7k\Omega$, and the temperature at which it
is measured, the only free fitting parameter is $I_C(T)$ yielding
4.8, 4.8, and 4.6\,nA respectively. Eq.\,\eqref{eq:1} fits the
measured $IV_J$ curves very well for all temperatures with $I_C$ as
the only fitting parameter. Above $\sim 300$\,mK smaller and smaller
critical currents are needed to make a good fit. Critical currents
versus temperature found by these fits are plotted in
Fig.\,\ref{fig:4}(a). At temperatures lower than $\sim 300$\,mK the
critical current is saturated at $\sim 5$\,nA, while at higher
temperatures it decreases more rapid than a BCS-gap dependence. In
Fig.\,\ref{fig:4}(a) we also plot the excess current versus
temperature, measured at $V_{sd} = 4\Delta_0/e$. We compare the
measurement with theory for a superconducting quantum point contact
\cite{martinrodero,shumeiko,cuevas,jorgensen}. We use Eq.\,1 and 2
in Ref.[\onlinecite{jorgensen}] with $\Delta=\Delta(T)$ having a BCS
temperature dependence to fit the measured temperature dependence of
the critical and excess current, solid red and blue curve in
Fig.\,\ref{fig:4}(a). The magnitude of the measured critical and
excess current is 0.25 and 0.7 lower than the theory predicts, while
their qualitative dependence on temperature fits well with theory.
\newline \indent
In Fig.\,\ref{fig:4}(b) we plot the magnetic field dependence of the
critical current on and off resonance (see arrows in
Fig.\,\ref{fig:1}(b)), and excess current on resonance. The critical
currents are found by the same method as above by fitting
Eq.\,\eqref{eq:1} to each measured $IV_J$ curve in
Fig.\,\ref{fig:2}(d). We compare the measurement to the same theory
as above, but with $\Delta = \Delta(B) = (1-B/B_C)\Delta_0$, where
$B_C \sim 90$\,mT is the critical field. We use a linear dependence
because, as shown in the insert of Fig.\,\ref{fig:4}(b), the sub-gap
structure has approximately a linear dependence on magnetic field.
The theory seems to fit qualitatively well to the measurement. But
the magnitude of the measured critical and excess current is, as
above for the temperature dependence, 0.25 and 0.7 lower than
theory.
\newline \indent
The dot is in the open regime with a charging energy of $U_C \sim
0.5$\,meV as discussed in the beginning of the paper, which is
several times larger than the superconducting energy gap ($\Delta_0
\sim 0.11$\,meV). We speculate that the discrepancy of the factor
0.25 between the measured critical current and theory could be due
to the charging energy being larger than the gap thus suppressing
the Cooper pair transport. The 0.7 discrepancy for the excess
current has been seen before \cite{jorgensen}, but we have no good
explanation for that.
\appendix
\section{Mean field description of Fabry-Perot resonances in a nanotube quantum dot.}
\begin{figure}
\begin{center}
\includegraphics[width=0.45\textwidth]{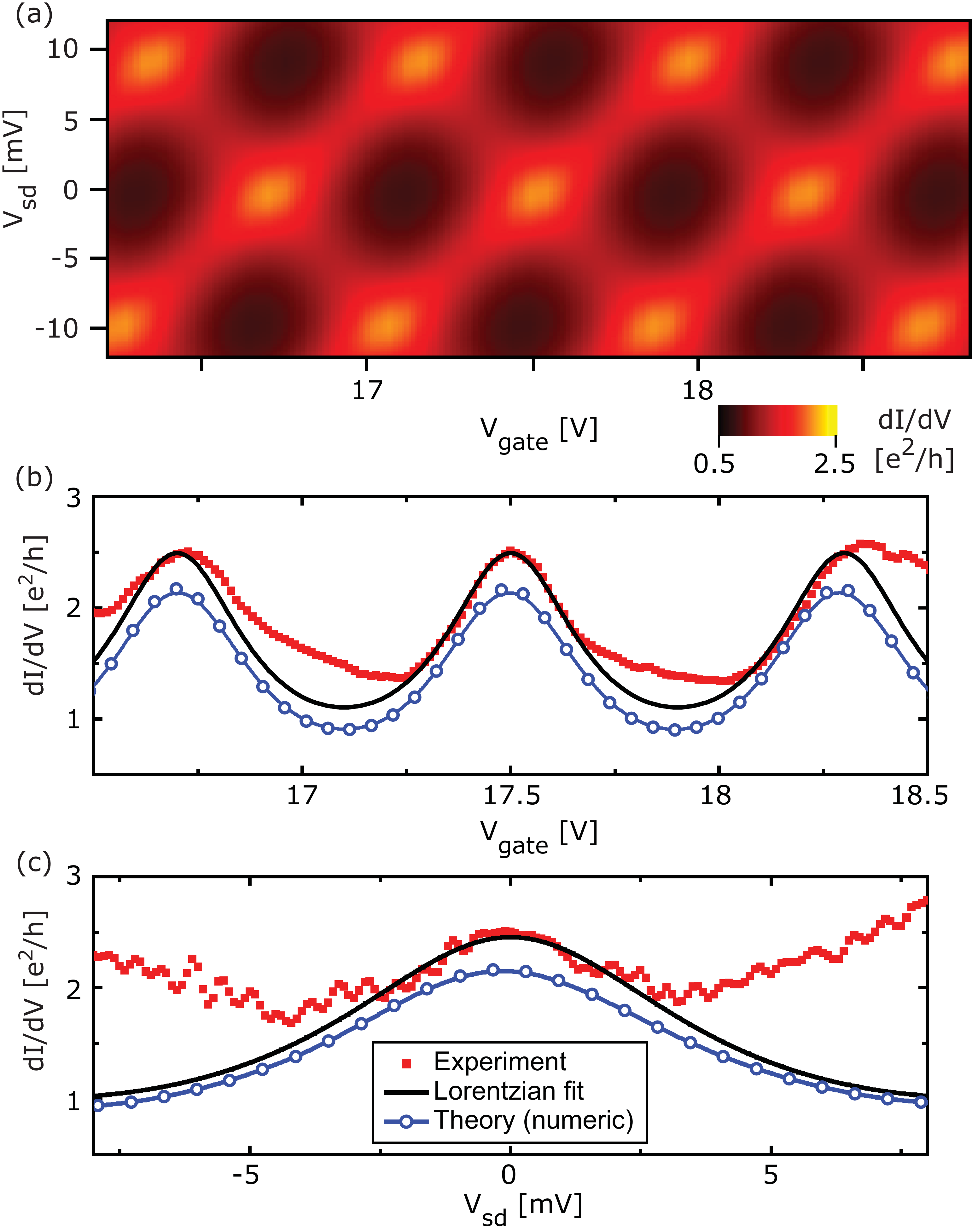}
\end{center}
\caption{(color online) (a) Differential conductance versus bias and
gate voltage using Eq.~\eqref{eq:N} and \eqref{eq:I}. (b) and (c)
Differential conductance versus gate voltage at $V_{sd}=0$\,mV (b),
and versus bias voltage at resonance (c). Red squares are
experimental data (measured with $B=150$\,mT), Solid black line is a
Lorentzian fit to the measurement yielding $W_g=0.4$\,V and
$W_{sd}=8.5$\,mV, and blue circles are numerical theory extracted
from (a).} \label{fig:app1}
\end{figure}
The electronic states in the nanotube can be described by
\begin{eqnarray}
H_{cnt}&=&\sum_{m\eta \sigma }\Delta E\, n_{m\eta \sigma
}+\frac12U_{C}\tilde{N}^{2}-e\,V_{eff}\,\tilde{N} \\
V_{eff}^{{}}&=&\sum_{\beta =g,s,d}\frac{%
V_{\beta }^{{}}C_{\beta }^{{}}}{C}
\end{eqnarray}
where
\begin{equation}
\tilde{N}=\sum_{m\eta \sigma }n_{m\eta \sigma },\quad
U_{C}=\frac{e^{2}}{C}
\end{equation}
and $\Delta E$ is the level spacing. The quantum numbers $m,\sigma
,\eta $ describe the orbital, spin and pseudospin degrees of
freedom, respectively. The subscripts g, s, and d refer to gate,
source and drain. In the experiment we apply asymmetric bias, i.e.,
$V_s = V_{sd}$ and $V_d=0$. In the mean-field approximation (which
is valid when $\Gamma \gg U_{C}$), the Hamiltonian is
\begin{equation}
H_{cnt}\approx \sum_{m\eta \sigma }\Delta E\, n_{m\eta \sigma
}+U_{C}\tilde{N}\langle \tilde{N}\rangle -eV_{eff}^{{}}\tilde{N},
\end{equation}
where the total occupation $\langle \tilde{N}\rangle $ should be
determined self-consistently
\begin{equation}
\langle \tilde{N}\rangle =\sum_{m\eta \sigma }\langle n_{m\eta
\sigma }\rangle ,
\end{equation}
with
\begin{equation}
\langle n_{m\eta \sigma }\rangle =\sum_{\alpha =s,d}\frac{\Gamma _{\alpha }}{%
\Gamma }\int \frac{d\omega }{2\pi }n_{F}(\omega +e\,V_{\alpha
})A_{m\eta \sigma }(\omega ).
\end{equation}
\begin{widetext}
Assuming all levels to be simple Lorentzians with equal widths, the
spectral functions are
\begin{equation}
A_{m\eta \sigma }(\omega )=\frac{\Gamma }{(\omega -m\,\Delta E
-U_{C}\langle \tilde{N}\rangle +e\,V_{eff}^{{}})^{2}+(\Gamma
/2)^{2}}.
\end{equation}
Inserting this into the integral, summing over quantum numbers, and setting $%
T=0,$ then gives the self-consistency equation
\begin{equation}\label{eq:N}
\langle \tilde{N}\rangle =\sum_{m}\sum_{\alpha =s,d}\frac{4\Gamma
_{\alpha }}{\Gamma }\left[ \frac{1}{2}-\frac{1}{\pi }\tan
^{-1}\left( \frac{m\,\Delta E +e\,V_{\alpha }+U_{C}\langle
\tilde{N}\rangle -e\,V_{eff}}{\Gamma /2}\right) \right] .
\end{equation}
This is equation can be solved numerically. Once we know the total
occupation for given gate, source and drain voltages, the current is
given by
\begin{eqnarray}\label{eq:I}
I &=&\frac{e}{h}\frac{4\Gamma _{s}\Gamma _{d}}{\Gamma }\sum_{m\eta
\sigma }\int \frac{d\omega }{2\pi }\left[ n_{F}^{{}}(\omega
+e\,V_{s})-n_{F}^{{}}(\omega +e\,V_{d})\right] A_{m\eta \sigma
}(\omega )
\nonumber \\
&=&\frac{4e}{h}\frac{4\Gamma _{s}\Gamma _{d}}{\pi \Gamma
}\sum_{m}\left[
\tan ^{-1}\left( \frac{m\,\Delta E +e\,V_{d}+U_{C}\langle \tilde{N}\rangle -e\,V_{eff}}{%
\Gamma /2}\right) \right.   \nonumber \\
&&\quad \left. -\tan ^{-1}\left( \frac{m\,\Delta E
+e\,V_{s}+U_{C}\langle \tilde{N}\rangle -e\,V_{eff}}{\Gamma
/2}\right) \right] .
\end{eqnarray}
\end{widetext}
In Fig.\,\ref{fig:app1}(a) we plot the differential conductance
versus bias and gate voltage using Eq.\,\eqref{eq:N} and
\eqref{eq:I} with the parameters found in
Eq.\,\eqref{eq:parameters}. We compare the theory with experimental
data measured with $=150$\,mT in Fig.\,\ref{fig:app1}(b) and (c). In
(b) we make a gate trace at zero bias and in (b) we make a bias
trace at the resonance indicated in Fig.\,\ref{fig:1}(a).
\end{document}